# Enhanced intrinsic voltage gain in artificially stacked bilayer CVD graphene field effect transistors

Himadri Pandey[a,b], Jorge Daniel Aguirre Morales[c], Satender Kataria[a,b], Sebastien Fregonese[c], Vikram Passi[b,d], Mario Iannazzo[e], Thomas Zimmer[c], Eduard Alarcon[e], Max C. Lemme[a,b,d]

[a] RWTH Aachen University, Chair for Electronic Devices, Otto-Blumenthal-Str. 2, 52074 Aachen, Germany

[b] University of Siegen, School of Science and Technology, Hölderlinstr. 3, 57076 Siegen, Germany

[c] IMS Laboratory, Centre National de la Recherche Scientifique, University of Bordeaux, Talence 33415, France

[d] AMO GmbH, Advanced Microelectronics Center Aachen, Otto-Blumenthal-Str. 25, 52074 Aachen, Germany

[e] Technical University of Catalonia, Department of Electronics Engineering, UPC, 08034 Barcelona, Spain

Corresponding author: max.lemme@uni-siegen.de

ABSTRACT

We report on electronic transport in dual-gate, artificially stacked bilayer graphene field effect transistors (BiGFETs) fabricated from large-area chemical vapor deposited (CVD) graphene. The devices show enhanced tendency to current saturation, which leads to reduced minimum output conductance values. This results in improved intrinsic voltage gain of the devices when compared to monolayer graphene FETs. We employ a physics based compact model originally developed for Bernal stacked bilayer graphene FETs (BSBGFETs) to explore the observed phenomenon. The improvement in current saturation may be attributed to increased charge carrier density in the channel and thus reduced saturation velocity due to carrier-carrier scattering.

INTRODUCTION

Graphene has attracted considerable attention from the electronic device engineering community due to its high charge carrier mobility and high saturation velocity [1] [2] [3]. Graphene field effect transistors (GFETs) have been demonstrated soon after the initial discovery of the material and it was established that GFETs made with single atomic layers of graphene suffer from high off-state leakage currents due to the absence of an energy band gap [4]. In addition, this absence of a band gap results in weak saturation in the device output characteristics (drain current versus source-drain voltage, $I_{DS}$-$V_{DS}$) [5]. This leads to a low maximum intrinsic voltage gain ($A_{v0}$) in monolayer GFETs, defined as the ratio of maximum direct current (DC) transconductance ($g_{m(max)}$) to minimum output conductance ($g_{d(min)}$) at the bias point where output conductance is minimum. $A_{v0}$ is typically reported to be less than or approaching one in monolayer GFETs [6] [7]. Nevertheless, a number of potential applications for gapless GFETs have been suggested. Experimentally demonstrated examples include frequency multipliers [8] or ambipolar radio frequency (RF) mixers [9], while other circuit applications have been proposed theoretically [10]–[13].

While some of these applications nevertheless appear to be promising, saturation is still desirable as it results in improved voltage gain and therefore improved device performance: $A_{v0}$ defines the maximum possible voltage gain when the device is suitably biased, which translates to the amplification performance of a GFET. In recent years, GFETs which utilize Bernal stacked bilayer (BSB) graphene as channel material have been reported to exhibit improved saturation in their output characteristics [7], [14]. This is because the presence of external vertical electric fields induce a small gate-tunable band gap of the order of a few hundred meV in Bernal stacked bilayer graphene [15]–[18]. Even though the performance of BSBGFETs is quite promising, the approach lacks in terms of scalability and applicability. Recently, there have been reports on large-area growth of bilayer graphene which may further enhance the potential of CVD bilayer graphene for various electronic applications [19]–[21]. However, one can not deny the very polycrystalline nature of CVD graphene and hence random stacking of grains in as-grown bilayer graphene. Liu et al. have demonstrated large-area CVD grown Bernal stacked bilayers [20], and the corresponding GFET performance of such material was found to be largely varying. Therefore, at present, availability of the material potentially puts a limit on the mass manufacturability and scalability of Bernal-stacked bilayer graphene FETs (BSBGFETs). Here, we propose the manual stacking of CVD monolayer graphene films to fabricate non-Bernal-stacked bilayer graphene FETs. The approach is promising in the sense that CVD monolayer growth is rather mature and one can obtain large-area graphene sheets on an industrial scale [22], as opposed to previous experiments with bilayer GFETs from folded monolayers [23]. We observe an enhanced tendency of current saturation in such artificially stacked (AS) bilayer GFETs (BiGFETs), leading to an improved $A_{v0}$. These results are discussed with the aid of Raman spectroscopy data of AS bilayer graphene sheets and electrical characterization data of BiGFETs carried out in ambient atmosphere. We also compare our data with a physics based compact model and Technology Computer Aided (TCAD) simulations, which substantiate our observations.

EXPERIMENT

  A.  DEVICE FABRICATION

Thermally oxidized (85 nm) silicon <100> wafers with a boron dopant concentration of $3\times10^{15}$ cm$^{-3}$ were diced and used as substrates. Bilayer graphene stacks were formed from two monolayer graphene sheets transferred onto the substrate with the help of a PMMA support layer. For this, CVD monolayer graphene (Moorfield NanoCVD) on copper foil was coated with 150 nm thick layers of poly methylmethacrylate (PMMA). Prior to transfer, the substrates were cleaned with acetone, iso-propyl alcohol (IPA) and di-ionized (DI) water. The graphene / PMMA films were then transferred onto the oxidized silicon substrates with an electro-chemical delamination method ("bubble method") [24]. The PMMA support layer was dissolved in acetone. Next, additional (i.e. second) monolayer graphene sheets were transferred onto the first layer to form artificially stacked bilayer CVD graphene. After removing the PMMA support layer on top of this stack, channels were patterned using photolithography and oxygen plasma assisted reactive ion etching (RIE). Source drain contacts were defined using optical lithography, followed by thermal evaporation of 20 nm chromium and 80 nm gold films and a lift-off process in acetone. 10 nm thick electron beam evaporated SiO$_2$ formed top-gate dielectrics, while 100 nm thick thermally evaporated aluminium layers formed top-gate electrodes. Two similar sets of devices with 10.7 nm of equivalent oxide thickness (EOT) of atomic layer deposited (ALD) Al$_2$O$_3$ ($\kappa = 9.1$, $t_{ox} = 25$ nm) were fabricated simultaneously, one with artificially stacked bilayer channels and the other with monolayer graphene channels. A global back gate was formed in all three cases by hydrofluoric acid (HF) assisted back-side oxide etch, followed by thermal evaporation of 300 nm thick gold layers. A schematic cross sectional view of a bilayer GFET is shown in **Figure 1** (a) whereas **Figure 1** (b) shows the top view optical micrograph of a fabricated device.

## B. RAMAN SPECTROSCOPY

The crystal structure and the corresponding band diagrams of monolayer and Bernal-stacked bilayer graphene are well documented. Likewise, the Raman fingerprint of these two materials have been established [25]. However, in the case of randomly stacked bilayer graphene sheets such as discussed in this work, the crystal structure, in particular how the layers are aligned with respect to one another, is not well defined, and so is also the band structure of such material. **Figure 2** (a) shows rough schematics of monolayer, Bernal stacked bilayer as well as randomly stacked AS bilayer graphene. Therefore, detailed Raman spectroscopy studies of the prepared bilayer graphene films were carried out (**Figure 2**). The 2D peak of artificially stacked bilayer graphene remains symmetric, similar to that of monolayer graphene. No clear splitting of the 2D band is observed in AS bilayer, which in contrast to the case of Bernal stacked bilayer, where an asymmetrical 2D band can be well fitted using four Lorentzian peaks [25]. However, a slight peak broadening (with a peak width of approximately 35 cm$^{-1}$) was observed in BiGFETs, accompanied by a small blue-shift as compared to monolayer graphene (here: 2D peak width of approximately 30 cm$^{-1}$). Compared to both monolayer and Bernal stacked bilayer graphene, the G peak intensity increased significantly. This has been demonstrated for twisted bilayer graphene to be twist angle dependent for twist angles between 3° and 27° between the two layers [26]. However, area maps of G and 2D band intensities (shown in insets in **Figure 2** (b)) revealed a significant non-uniformity in their intensities, pointing towards random overlapping of grains in AS bilayer graphene. The D peak, which denotes the presence of defects in the graphene layer(s) did not exhibit any significant changes as compared to monolayer graphene. These observations are in line with previous reports, which attribute them to the presence of weak electronic coupling between the two artificially stacked layers [26]. Furthermore, a pronounced blue shift in Raman spectra in bilayer graphene systems may also result from changes in Fermi velocity at low twist angles between two grains. Recently, tunable band gap (i.e., electrical properties modification/enhancement) has also been demonstrated in CVD synthesized twisted bilayer graphene [27], and the same has been attributed to the combined effects of applied vertical fields as well as twist angles. Therefore, one may expect that a weak electronic interaction between two graphene layers (not necessarily Bernal stacked) can also influence the performance of GFETs. .

RESULTS & DISCUSSIONS

All devices investigated in this work were characterized in ambient atmosphere conditions with a Keithley 4200 SCS Semiconductor parameter analyzer connected to a Karl Süss probe station. **Figure 3** (a) and (b) show the output and transfer characteristics of a BiGFET at zero back-gate voltage ($V_{BG}$). The device shows the typical features of graphene FETs, including ambipolar transfer curves ($I_{DS}$-$V_{TG}$) and quasi-saturating output characteristics ($I_{DS}$-$V_{DS}$). In all the measurements reported here, the top and back gate leakage currents were observed to be between a few tens of pico-amperes to a few tens of nano-amperes. These leakage currents are low enough for gate oxides to be considered as 'working'. Contact resistance values varied between 18 kΩ to 27 kΩ, which were extracted using a model proposed in [28]. We note that these devices were not optimized with regards to their contact resistance values. However, doing so would certainly further improve the observed electrical properties [29]. The level of current saturation in these devices is found to be enhanced compared to monolayer GFETs at the $V_{BG}$ = 0 V condition. From these basic measurements, key figures of merit can be extracted such as the transconductance $g_m$, and the output conductance $g_d$.

We investigated these key parameters further at non-zero $V_{BG}$ conditions. Transfer characteristics were measured at various drain biases ($V_{DS}$) and output characteristics were measured at various top-gate voltages ($V_{TG}$) at different fixed $V_{BG}$ (totaling over 100 curves, example plots shown in **Figure 3** (c) and (d)).

**Figure 3** (e) and (f) show the maximum DC transconductance values $g_{m(max)}$ and minimum output conductance values $g_{d(min)}$ plotted as a function of $V_{BG}$ for BiGFETs and CVD monolayer GFETs, respectively. $g_{m(max)}$ is defined as the maximum value of DC transconductance ($g_m$) obtained at a given $V_{BG}$ when various $g_m$ vs. $V_{TG}$ curves are measured, one at a time for a specific $V_{DS}$ value. Likewise, $g_{d(min)}$ is defined as the minimum value of output conductance ($g_d$) obtained at a given $V_{BG}$ when various $g_d - V_{DS}$ curves are measured, one at a time for a specific $V_{TG}$. In these data, several observations are important for the maximum intrinsic voltage gain $A_{v0}$: When compared to monolayer GFETs, BiGFETs have slightly higher variations in $g_{m(max)}$ as shown in **Figure 3** (e) and generally have lower $g_{d(min)}$ as shown in **Figure 3** (f). First, we will discuss the magnitude and mechanism for (non) gate dependence of $g_{m(max)}$. The application of two opposite polarities (here: positive $V_{TG}$ and large negative $V_{BG}$) results in the formation of a dominantly n-channel under the gate and a dominantly p-region in the ungated access, respectively. This results in the formation of a p-n junction close to the drain terminal, as shown schematically in **Figure 4** (a). In monolayer graphene, the absence of a band gap results in a nearly transparent p-n junction due to Klein tunneling. In the case of BiGFETs, the peak p-n junction resistance $R_{pn}$ – extracted following the method described in [30] – is found to be of similar magnitude as that in a monolayer GFET without a band gap (**Figure 4** (b)). In the case of Bernal stacked bilayer GFETs, where the band gap depends on the back gate voltage, the opening of the band gap increases the p-n junction resistance and hence, the device series resistance and $g_{m(max)}$ [7], [31]. The latter is very apparent in a comparison of extracted BiGFET and monolayer GFET data with Bernal-stacked bilayer GFETs (from literature [7], **Figure 5** (a)). This leads to the following conclusions: (1) BiGFETs do not show signs of an electronic band gap through electrical data. (2) The observed behavior corroborates Raman data, which show that two stacked monolayers are largely electronically independent.

Similar to $g_{m(max)}$, the extracted values for $g_{d,min}$ do not show gate-dependence in both BiGFETs and monolayer GFETs (**Figure 3** (f)). However, we observe that $g_{d(min)}$ is lower in BiGFETs than in monolayer GFETs, and in fact is similar to that in BSBGFETs (literature data, **Figure 5** (b)). The extracted $g_{d(min)}$ values are in the range of 0.002 to 0.015 mS/μm (BIGFETs) and 0.002 to 0.004 mS/μm (BSBGFETs are between, [7]), as opposed to 0.02 to 0.05 mS/μm (monolayer GFETs). As reasoned in the first part of this work, BiGFETs do not show signs of a energy band gap, which is known to be the responsible factor for low $g_{d(min)}$ in BSBGFETs [31]. Nevertheless, an effective increase in intrinsic voltage gain $A_v$ (=$g_m/g_d$) is observed compared to monolayer GFETs. $A_{v0}$ (=$g_{m(max)}/g_{d(min)}$) in BiGFETs was found to lie between 2.5 and ~29, which is between monolayer GFETs and BSBGFETs [7]. **Table 1** summarizes these figures in comparison to literature data. While this observation leads to the preliminary conclusion that BiGFETs may be favorable for analog applications over monolayer GFETs, we will conclude this work with a possible explanation for the behavior based on a device model.

First, the induced charge carrier density in the channel was calculated using a mathematical model proposed in [32]. According to this model, the induced carrier density in the channel of a monolayer GFET under the influence of applied gate voltages is given by

$$n_{total} = \sqrt[2]{n_0^2 + (1/q)^2 * \{C_{top}(V_{TG} - V_{DiracTG}) + C_{back}(V_{BG} - V_{DiracBG})\}^2} \quad (1)$$

where, $n_{total}$ defines total induced carrier density in the channel, $n_0$ is the residual carrier density in graphene, $C_{top}$ & $C_{back}$ represent top-gate & back-gate capacitance, $q$ denotes elementary charge, $V_{TG}$ is top-gate voltage, $V_{DiracTG}$ and $V_{DiracBG}$ denote the corresponding top-gate & back-gate Dirac point locations. Under the assumption that the two stacked graphene monolayers behave as one (artificial) bilayer in the channel, the above equation was applied to our BiGFET system to calculate carrier densities in the channel at various $V_{DS}$ and $V_{BG}$ combinations. The value of $n_0$ was adopted to be $1\times10^{12}$ cm$^{-2}$ assuming a relatively dirty, and thus more realistic, GFET system in accordance to the criterion suggested in literature [33]. A monolayer GFET fabricated alongside these devices with similar material parameters was also characterized and induced carrier densities were assessed for a comparison. **Figure 6** (a) shows the trends extracted for these devices for two different dielectrics i.e. Al$_2$O$_3$ and SiO$_2$. It can be seen clearly that, as the source-drain bias and hence the lateral electrical field in the device increased even at $V_{BG}$ = 0 V condition, the carrier population increases at a steeper rate in the BiGFET channel than in the monolayer GFET channel, irrespective of the dielectric type [34]. This can be understood when considering that the carrier population is assessed with respect to the location of the Dirac voltage ($V_{DiracTG}$) in the transfer curve, using Eq. (1), at a fixed source-drain bias. This drain induced Dirac shift (DIDS) [35] is higher in BiGFETs, always larger than 0.5V with every 1V change in $V_{DS}$ irrespective of the dielectric and back-gate voltage conditions. In monolayer GFETs, DIDS was always less than 0.5V under similar conditions (**Figure 6** (b)). However, it should be noted that BiGFETs with Al$_2$O$_3$ dielectric exhibited better carrier density modulation compared to SiO$_2$. The difference in performance of BiGFETs based on different dielectrics, Al$_2$O$_3$ and SiO$_2$, may be prima facie attributed to the difference in their qualities and dielectric constants. Since Al$_2$O$_3$ is ALD deposited from a very thin (~2 nm) Al seed layer, its interface quality & uniformity is better compared to e-beam evaporated SiO$_2$, which is more prone to having high defect densities. A detailed analysis on dielectric properties and their influence on device behavior is beyond the scope of this paper, but we refer to separate investigations on this matter [36]–[38].

This observation attests that the carrier population in BiGFETs grows relatively higher than that observed in monolayer GFETs under similar biasing conditions. When viewed under the classical drift-diffusion transport approach, which applies for our micrometre sized devices, we expect that a carrier population increase in a device will reduce the drift velocity of the carriers due to carrier-carrier scattering in the two-dimensional transport plane. This can be understood with the equation (2),

$$I_{SD} \propto \int_0^{L_g} n(x) * V_{Drift}(x) dx, \qquad (2)$$

with the carrier population dependent saturation velocity in the channel correlated as,

$$v_{sat} = \Omega \Big/ \sqrt{(\pi * n_{total})} \qquad (3)$$

where $I_{SD}$ represents source-drain current, $n(x)$ defines carrier population at location $x$ away from the drain end, $V_{drift}(x)$ defines the corresponding carrier drift velocity and $\Omega$ represents the substrate phonon interaction energy. Therefore, this implies that the BiGFET channels experience more carrier-carrier scattering per unit applied field gradient as lateral and vertical electrical fields increase, than monolayer GFET channels. Therefore, BiGFETs would be expected to show reduced saturation velocity, which explains the enhanced tendency of saturation in BiGFET channels and the reduced $g_{d(min)}$ values in **Figure 3** and **Figure 5**. Note that the substrate in all the devices remains unchanged, and therefore $\Omega$ in this case should be constant.

In order to further validate the assumptions made to explain the observed device characteristics, the experimental data was fitted using an established compact model described in [28]. The model is developed in Verilog-A language and is validated against experimental data from BSBGFETs with electrically tunable band gap published in literature. However, as we have established that the bandgap in BiGFETs should be insignificant, the corresponding parameter in the compact model has been set to 0. **Figure 7** shows transfer and output characteristics obtained from one of our BiGFETs fitted to this model by tuning the parameters accordingly within the limits of justification. A very good fit with the average root mean square (RMS) error value always lying below 10% was obtained for $V_{DS}$ values between 0 to +3.5V, $V_{TG}$ values between -5 to +5V and several $V_{BG}$ values between -60V & +60V. At this point, the extracted induced carrier density in the center of the channel was found to be between -9.8x10$^{12}$ cm$^{-2}$ (electrons) & 1.3x10$^{13}$ cm$^{-2}$ (holes), respectively. This is qualitatively of same order of magnitude as that calculated using the model described in [32] and plotted in **Figure 6**. The extracted electron & hole mobility values in the channel region were found to be 830 cm$^2$/V.s and 610 cm$^2$/V.s, respectively. The electron-hole mobility asymmetry in GFETs is well reported and can be attributed to factors like contact induced doping, field induced doping or phonon-induced scattering [32], [39] [40]. The close fit with experimental data in **Figure 7** could only be obtained when the factors describing the Dirac point shift due to back-gate voltage are reduced to a non-physical value, because compared to a Bernal stacked bilayer case, the Dirac point shift in our devices is much smaller. It was found that the expected range of Dirac shift is (0, 15V) assuming an ideal BSBGFET, whereas the measured range was only (0, 600mV). This shows that the effect of back-gate on the location of the Dirac voltage is very small, as is also observed in **Figure 3** (c). We have identified several potential reasons for this very weak dependence of the Dirac voltage shift due to back-gate voltage:

1: The value of the back-gate oxide thickness is not correct; this reason has been excluded because the $t_{ox}$ has been measured by ellipsometry to be 85 nm.

2: The presence of high back-gate leakage currents would undermine the effect of back-gate voltage; the measured leakage currents were limited to a few pA (data not shown here), so this possibility was dismissed.

3: The presence of a depletion region under the back-gate would effectively shield the effect of back-gate voltage on the channel.

This latter possibility was explored with TCAD simulations of BiGFETs using the cross sectional schematic depicted in **Figure 1** (a). It was found that indeed an inversion/accumulation/depletion region is formed under the back-gate oxide interface, with a thickness between a few hundred nanometers up to a few micrometers as the applied $V_{BG}$ is varied. **Figure 8** (a) summarizes the results of the TCAD simulations for a several substrate doping concentrations (including $3 \times 10^{15} cm^{-3}$, as used in the experimental part of this work) and resulting equivalent capacitances that are formed due to the inversion/accumulation/depletion region under the channel. Moreover, as the carrier concentration in the bulk is varied with different applied back-gate voltages, the depth of depletion region into the bulk of the substrate also varies, as shown in **Figure 8** (b). This inversion/accumulation/depletion region results in an additional capacitance element and leads to a much reduced back-gate voltage induced Dirac voltage shift in our devices.

The extracted carrier mobility values in BiGFETs are somewhat lower than in similar monolayer GFETs reported in literature [4] [7] [32][41]. The mobility is thought to be limited because each of the artificially stacked monolayers is CVD grown and - despite low defect densities as indicated in the Raman scan (**Figure 2**) - inevitably includes grain boundaries. These act as scattering centres for carriers, thus lowering the effective mobility values [42]. More importantly for the understanding of our results, the high carrier density observed in the channels results in increased carrier-carrier scattering at higher voltages, and hence increased saturation. The physics based model thus confirms our assumption that a higher carrier density in BiGFET channels leads to the observed enhanced saturation tendency in the output characteristics of these devices.

CONCLUSIONS

In this work, artificially stacked large-area CVD graphene field effect transistors (BiGFETs) have been investigated under ambient conditions. Detailed Raman spectroscopy has been carried out to understand the nature of the artificially stacked graphene bilayer, which revealed weak electronic coupling between the two stacked monolayers. The electric transport in these devices has been studied at high bias conditions using top- and bottom gates. Key analog performance parameters have been compared with monolayer GFETs and Bernal stacked bilayer GFETs. An enhanced tendency to current saturation is observed in the AS bilayer graphene devices. The experimental data thus obtained has been used to propose a physics based compact model, originally developed for describing electrical transport in Bernal stacked bilayer graphene FETs. A very good fit with low (<10%) RMS error has been obtained in a wide range of gate voltages and source drain bias values if the electronic band gap in the model is set to 0 eV, in accordance with our experimental data. The carrier density extracted from the model in the center of the device channel is of the order of ~$10^{13}$ cm$^{-2}$. Thus, the enhanced tendency to saturation in the output characteristics observed in these devices may be attributed to carrier-carrier scattering occurring under high bias measurement conditions., however, other factors such as phonon induced scattering [32], Joule heating [43] or scattering at defects induced by residual PMMA may also play a role [44]. This enhanced saturation leads to an improved intrinsic voltage gain in BiGFETs. We extract in the DC regime favourable high frequency performance figures of merit of BiGFETs, namely $g_{m(max)}$ (which proportionately affects maximum cut off frequency $f_T$) and improved $g_{d(min)}$ (which enhances maximum oscillation frequency $f_{max}$). The devices remain fully functional even at high bias conditions. Therefore, such artificially stacked large-area CVD bilayer graphene devices may offer a scalable alternative solution for future radio frequency and analog circuit applications.

ACKNOWLEDGMENT

This work was supported by the German Research Foundation (DFG, LE 2440/1-1 & 2-1), by the European Commission (GRADE, 317839 and ERC grant InteGraDe, 307311) and by the Spanish Ministry of Science & Innovation under projects RUE CSD2009-00046 and TEC2010-15765.

| Device type | Maximum intrinsic gain $A_{v0}=g_{m,max}/g_{d,min}$ |
|---|---|
| **Monolayer GFET** | 6 [7] |
| **BiGFET** | Between 2.5 to 29 |
| **BSBGFET** | 35 [7] |

**Table 1:** Intrinsic voltage gain in BiGFETs compared with literature data in [7]. The improvement is mainly attributed to reduction in $g_{d(min)}$ values in BiGFETs.

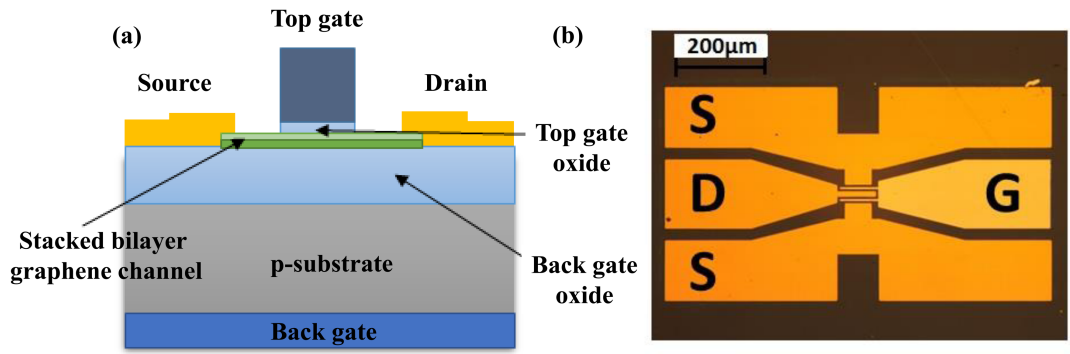

**Figure 1:** (a) Cross sectional schematic of a stacked bilayer graphene FET. (b) Optical micrograph of a completed device after final lift-off step. S, D & G represent source, drain and gate terminals respectively.

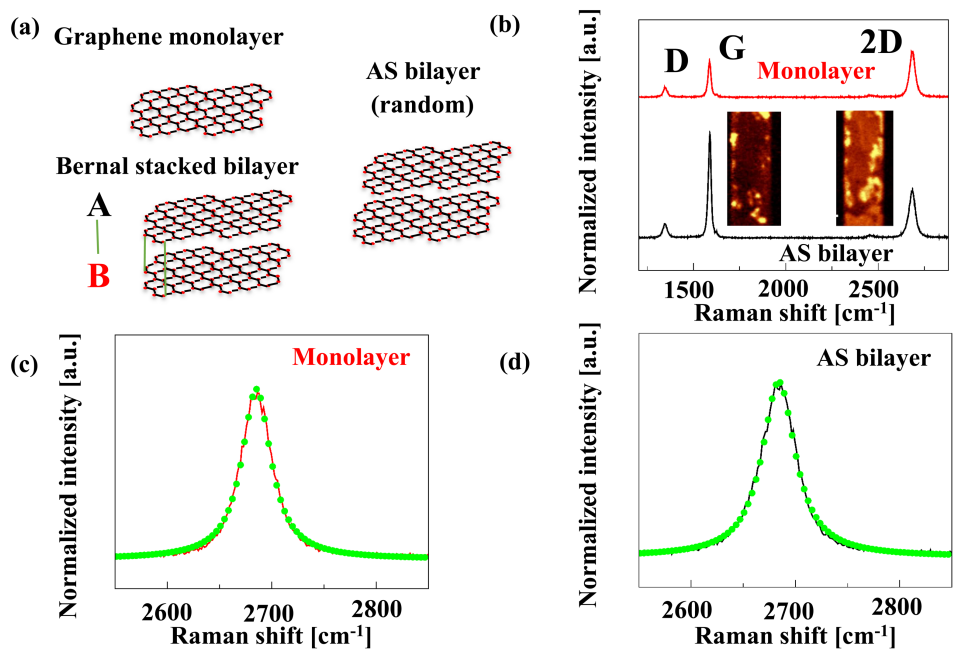

**Figure 2:** (a) Schematics of monolayer, Bernal stacked bilayer and artificially stacked (AS) bilayer graphene. Two adjacent lattice positions A & B in each of the monolayers are outlined. In Bernal stacked bilayers, every A atom of the top layer is aligned and van der Waals bonded to every B atom, resulting in a particular crystal structure. In AS bilayer graphene, randomly oriented grains in the first layer are stacked on to randomly oriented grains in the second layer, at random angles. (b) Raman spectra measured on monolayer graphene and AS bilayer graphene for which no clear information about the crystal alignment or the band structure is available. The inset shows the area maps of G (left side) and 2D peak (right side) intensities. A significant variation in the respective intensities indicates random overlapping of grains in AS bilayer graphene. Zoomed-in view of the 2D peak of (c) monolayer and (d) AS bilayer graphene. The green dotted curves in (c) and (d) indicate the corresponding Lorentzian fits, outlining the observation that only one Lorentzian fit describes well not only the 2D peak of monolayer graphene but also that of AS bilayer graphene. Even though slight peak broadening is observed in case of AS bilayers, this signature hints at two rather independent layers of graphene that are electronically different from their Bernal stacked bilayer counterparts.

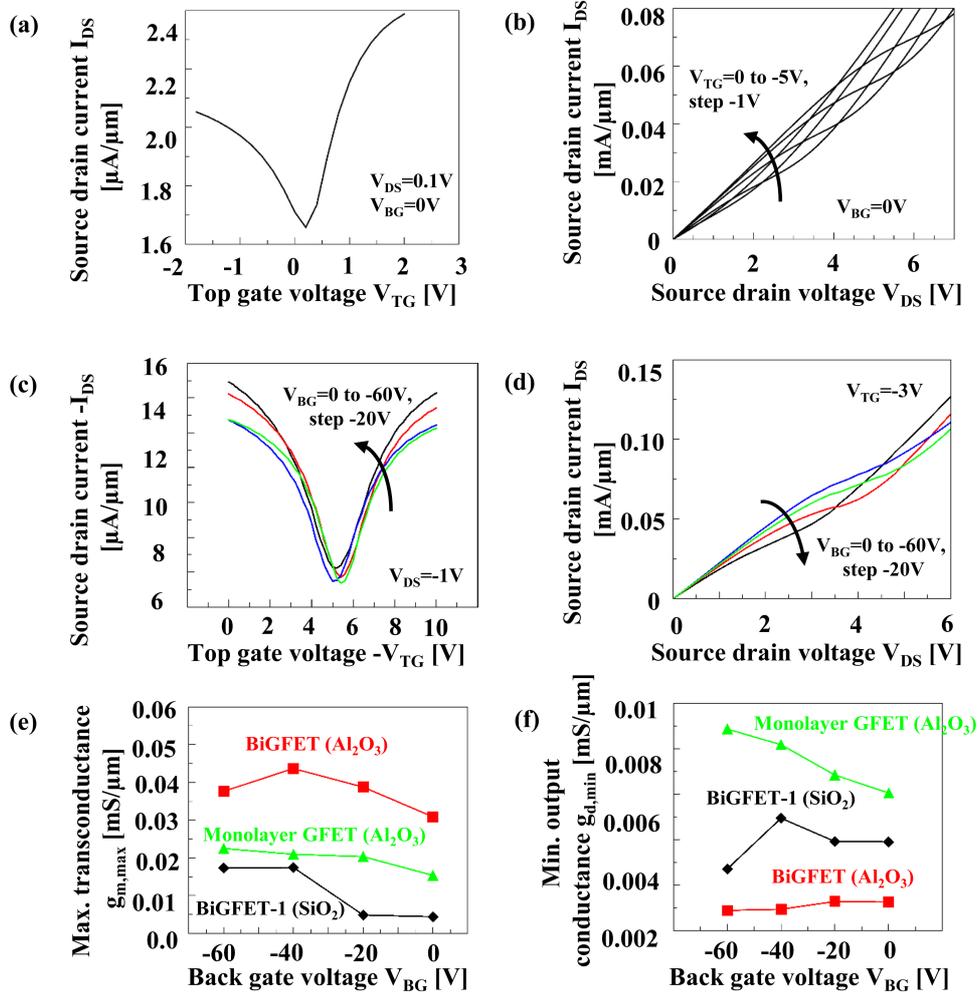

**Figure 3:** (a) Transfer characteristics ($I_{DS}$-$V_{TG}$) and (b) output characteristics ($I_{DS}$-$V_{DS}$) of a BiGFET with a gate length of $L_g$ = 4 μm and a channel width of W = 40 μm. (c) transfer and (d) output characteristics of BiGFETs are shown at various back gate voltages (0 to -60V, step -20V). At each back gate voltage, several such characteristics were measured and corresponding DC transconductance $g_m$ and output conductance $g_d$ were then extracted. (e) Maximum DC transconductance $g_{m(max)}$ and (f) minimum output conductance $g_{d(min)}$ values plotted as a function of back gate voltage for various types of BiGFETs, compared to CVD monolayer GFETs. Almost no tunability is observable for $g_{m(max)}$, and $g_{d(min)}$ is generally lower than that of monolayer GFETs. All devices in (c), (d), (e) and (f) had $L_g$=12 μm and W = 60 μm.

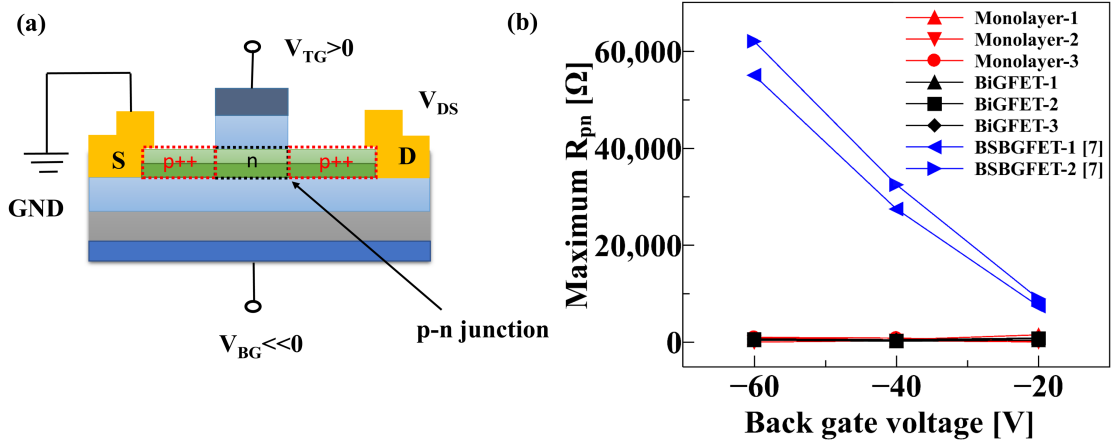

**Figure 4:** (a) Schematic showing the formation of a p-n junction in BiGFET channels due to a large back gate voltage. (b) Maximum p-n junction resistance $R_{pn}$ as a function of back gate voltage for monolayer GFETs and BiGFETs (devices shown in **Figure 3**) and BSBGFETs (literature data, [31]). The magnitude of $R_{pn}$ is similar in BiGFETs and in monolayer GFETs, while BSBGFETs show much higher values due to a semiconducting p-n junction.

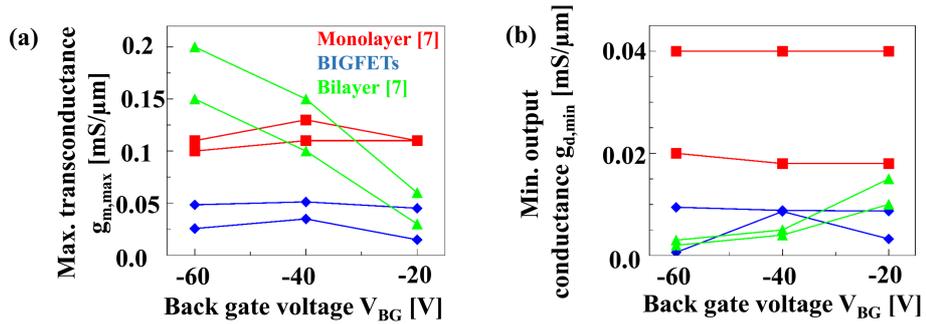

**Figure 5**: (a) Maximum DC transconductance ($g_{m(max)}$) and (b) minimum output conductance ($g_{d(min)}$) values of BiGFETs plotted as a function of back-gate voltage values. The data points of the red and green symbols are taken from [7]. Two devices from each type are plotted for comparison. For BiGFETs, $g_{m(max)}$ values in show a negligible back gate-tunable behaviour, much weaker than that observed in BSBGFETs. Here it has to be noted that $g_{d(min)}$ values of BiGFETs are usually lower than those of monolayer GFETs, however no clear trend is observed, as concluded from (b). All BiGFET devices shown above have L$_g$=12 μm and W=60 μm.

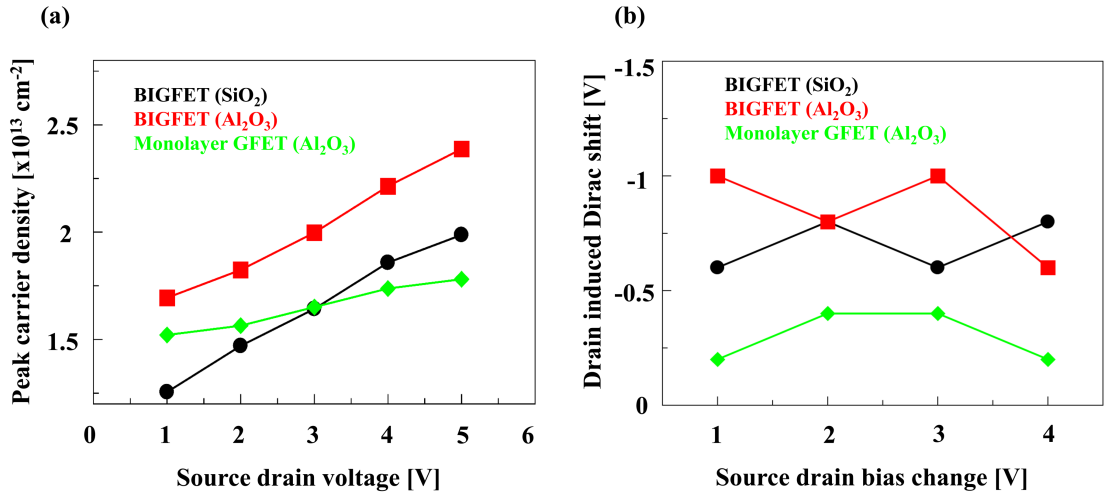

**Figure 6:** (a) Peak carrier density values calculated using the model described in [32] at various source-drain bias voltages and under $V_{BG}$ = 0 V condition. BiGFETs show enhanced induced carrier densities in the channel irrespective of dielectric type, whereas the monolayer GFET shows less steep growth in carrier population even at higher bias conditions. (b) Drain induced dirac shift (DIDS) for the devices in (a) with respect to applied source drain bias values at $V_{BG}$ = 0 V. DIDS for BiGFETs was always observed to be more than 0.5 V per 1 V change in $V_{DS}$. Gate length was 3 μm for both devices with $Al_2O_3$ gate dielectric and 4 μm for the devices with $SiO_2$ gate dielectric. Channel width in all cases was 40 μm.

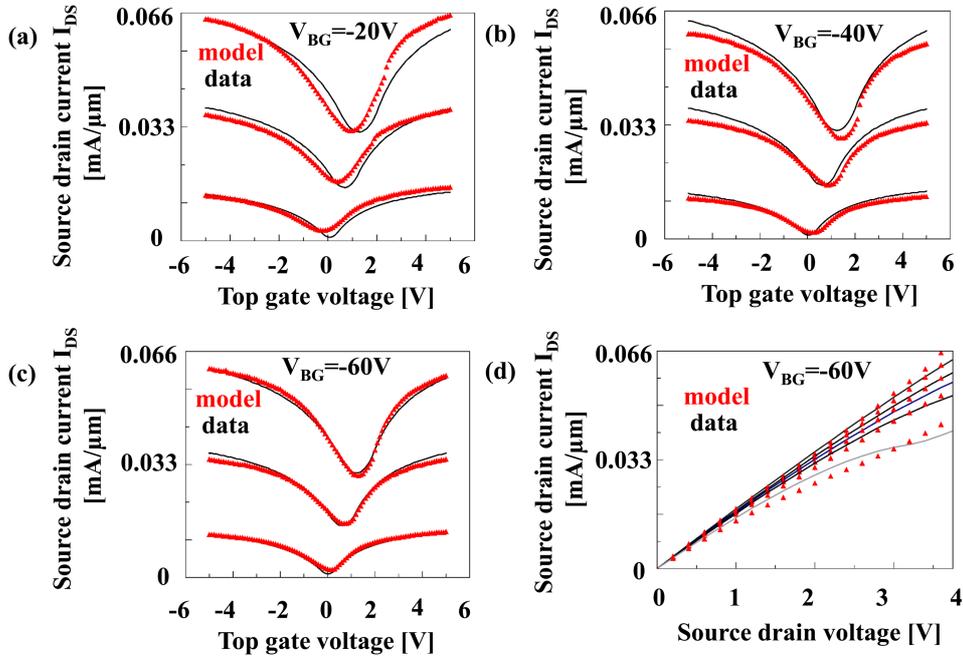

**Figure 7:** Experimental data from a BiGFET fitted against a compact model developed for Bernal stacked bilayer GFETs [28] at various back-gate voltage conditions. The value for the band gap in the model has been set to 0, based on the experimental results of this work. (a), (b)

and (c) show transfer characteristics obtained for $V_{DS}$=1,2 and 3V. (d) shows output characteristics between $V_{TG}$=0 to 5V. The BiGFET device had a gate length $L_g$=12µm and channel width W=60µm. As can be seen directly from the model based fitting, a good fit is obtained with modest root mean square error (<10% in all cases) within the voltage ranges adopted during measurements at various back-gate voltages. Black lines depict experimental data while red symbols indicate simulation data.

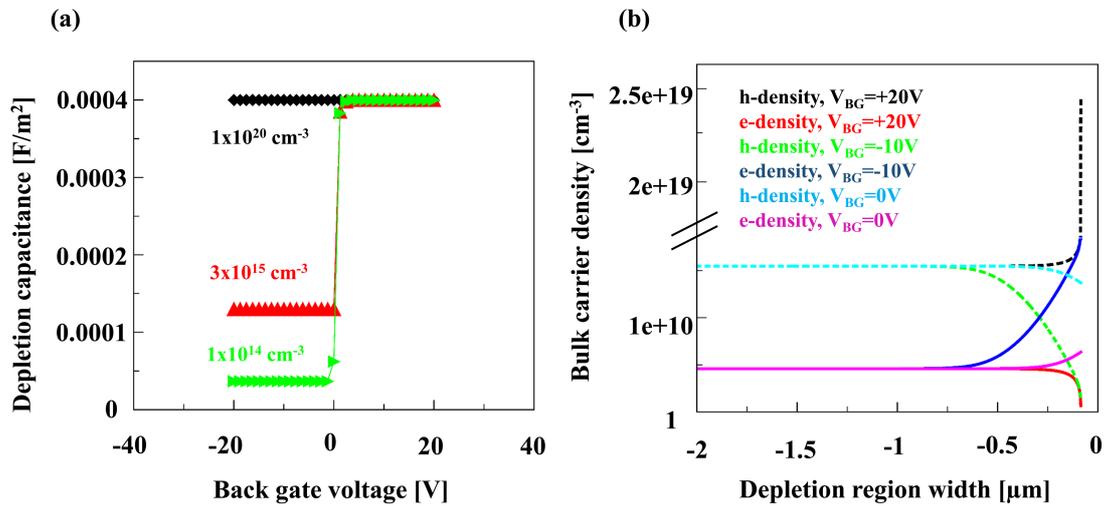

**Figure 8:** TCAD simulation: (a) formation of a depletion capacitance in presence of external back-gate voltages for a range of dopant concentrations. (b) Formation of depletion region up to a certain depth by carrier density variations occurring in the bulk silicon substrate (for selected values of $V_{BG}$). These results together explain the limited Dirac voltage shift observed in the experiments compared to the modified BSBGFET compact model.